\documentclass{aastex}
\usepackage{emulateapj5,apjfonts,psfig}

\makeatletter

\newenvironment{figurehere}
  {\def\@captype{figure}}
  {}
\makeatother

\def\etal{{et al.\ }}

\def\cm{{\rm\thinspace cm}}

\def\erg{{\rm\thinspace erg}}

\def\keV{{\rm\thinspace keV}}
\def\km{{\rm\thinspace km}}

\def\Lsun{\hbox{$\rm\thinspace L_{\odot}$}}

\def\Msun{\hbox{$\rm\thinspace M_{\odot}$}}

\def\ph{{\rm\thinspace ph}}
\def\s{{\rm\thinspace s}}

\def\ergps{\hbox{$\erg\s^{-1}\,$}}

\def\kmps{\hbox{$\km\s^{-1}\,$}}

\def\powerlawfluxat1kev{\hbox{$\ph\cm^{-2}\s^{-1}\keV^{-1}$}}
\def\ew{$W_{\rm K\alpha} \;$}

\def\lapp{\ifmmode\stackrel{<}{_{\sim}}\else$\stackrel{<}{_{\sim}}$\fi}

\def\gapp{\ifmmode\stackrel{>}{_{\sim}}\else$\stackrel{>}{_{\sim}}$\fi}
\def\spose#1{\hbox to 0pt{#1\hss}}

\def\approxlt{\mathrel{\spose{\lower 3pt\hbox{$\sim$}}
        \raise 2.0pt\hbox{$<$}}}
\def\approxgt{\mathrel{\spose{\lower 3pt\hbox{$\sim$}}
        \raise 2.0pt\hbox{$>$}}}

\def\lapp{\ifmmode\stackrel{<}{_{\sim}}\else$\stackrel{<}{_{\sim}}$\fi}
\def\gapp{\ifmmode\stackrel{>}{_{\sim}}\else$\stackrel{>}{_{\sim}}$\fi}

\def\mcg6{MCG$-$6-30-15}
\def\mr2251{MRC~2251-178}
\def\ngc2110{NGC~2110}
\def\iras13349{IRAS~13349+2438}
\def\iras18325{IRAS~18325--5926}
\def\grs1915{GRS~1915+105}
\def\xtej1748{XTE~J1748-288}
\def\chandra{{\it Chandra }}
\def\XMM{{\it XMM-Newton }}
\def\xmm{{\it XMM }}
\def\rxte{{\it RXTE }}
\def\asca{{\it ASCA }}

\def\feka{$F_{\rm K\alpha} \;$}

\def\xtegammamcg6{$\Gamma=1.9$}

\def\fe25{Fe~{\sc xxv}}
\def\fe26{Fe~{\sc xxvi}}
\def\Ne9{Ne~{\sc ix }}
\def\ne10{Ne~{\sc x }}
\def\mg11{Mg~{\sc xi }}
\def\si13{Si~{\sc xiii }}

\def\apj{ApJ}

\def\mnras{MNRAS}
\def\apj{ApJ}
\def\apjs{ApJS}
\def\aap{A\&A}

\def\c2{{\sc C~ii}}
\def\c3{{\sc C~iii}]}
\def\c4{{\sc C~iv}}
\def\n5{{\sc N~v}}
\def\o3{[{\sc O~iii}]}

\def\si4{Si~{\sc iv}}
\def\fe25{Fe~{\sc xxv}}
\def\fe26{Fe~{\sc xxvi}}
\def\mg2{Mg~{\sc ii}}
\def\Msun{\ifmmode M_{\odot} \else $M_{\odot}$\fi}
\def\Lsun{\ifmmode L_{\odot} \else $L_{\odot}$\fi}

\slugcomment{Accepted for publication in the  Astrophysical Journal Letters}

\shorttitle{The Chandra HETGS view of the Fe~K$\alpha$ emission in \mcg6 }

\shortauthors{LEE et al.}

\begin{document}

\title{The shape of the relativistic iron K$\alpha$ line from \mcg6 \\ measured with the Chandra HETGS and RXTE}
\author{
Julia C. Lee\altaffilmark{1},
Kazushi Iwasawa\altaffilmark{2},
John C. Houck\altaffilmark{1},
Andrew C. Fabian\altaffilmark{2},
Herman L. Marshall\altaffilmark{1},
Claude R. Canizares\altaffilmark{1}
 }
\altaffiltext{1}{MIT, Department of Physics and Center for Space Research, 77 Massachusetts Ave., NE80, Cambridge, MA 02139.}
\altaffiltext{2}{Institute of Astronomy, University of Cambridge, Madingley Rd., Cambridge CB3 0HA  U.K.}

\medskip
\centerline{To appear in {\sc The Astrophysical Journal Letters}}

\begin{abstract}
We confirm the detection of the relativistically broadened iron
K$\alpha$ emission at 6.4~keV with simultaneous \chandra HETGS and
\rxte PCA observations.  Heavily binned HETGS data show a disk line
profile with parameters very similar to those previously seen by
{\it ASCA}.  We observe a resolved narrow component with a velocity width
$\sim 4700 \, \rm km\,s^{-1}$ (FWHM $\sim 11,000 \rm km\,s^{-1}$),
that is most prominent, and narrower (FWHM $\sim 3600 \, \rm km \,
s^{-1}$)  when the continuum flux is high.  It plausibly is just the
blue  wing of the broad line. We obtain a stringent limit on the
equivalent width of an intrinsically narrow line in the source of
16~eV, indicating little or no contribution due to fluoresence from 
distant material such as the molecular torus.  Variability studies 
of the narrow component show a constant iron line flux and 
variable width indicating the line may be originating from 
different kinematic regions of the disk.

\end{abstract}


\section{Introduction} \label{sec:intro}
The discovery with \asca of broad iron lines in active galactic nuclei
(AGN; Tanaka et al. 1995, hereafter T95; Mushotzky et al. 1995; Nandra
et al. 1997), which were predicted by Fabian et al. (1989), has opened
the window for studying the  innermost regions of the accretion flow
and the strong gravity  regime at about 10 gravitational radii
(i.e. $10r_g=10GM/c^2$) around a black hole (Fabian et al. 1995).  At
the present time this regime is only accessible to X-ray observations
(optical and radio spectral line data from galactic nuclei probe
matter at $\sim 10^5 r_g$), and the iron line  is widely accepted to
be a comparatively robust probe of the immediate environment closest
to the supermassive black hole at the center of AGNs (see review by
Fabian et al. 2000, and references therein).

The best studied source and first to show the most extreme broad
($\Delta$v $\rm \sim \, 100,000 \, km \, s^{-1}$, T95) iron K$\alpha$
emission line is the luminous ($\rm L_X \sim 10^{43} \, \ergps$)
Seyfert~1 galaxy \mcg6 at $z \sim 0.0078$. The long 4.5-d {\it ASCA}
observation in 1994 which revealed the line to be broad also showed it
to be skewed (T95; Iwasawa et al. 1996). The breadth and asymmetry of
this line was subsequently confirmed by more long {\it ASCA} (Iwasawa
et al. 1999, Shih, Iwasawa \& Fabian 2002), {\it RXTE} (Lee et al.
1998, 1999)  and {\it BeppoSAX} (Guainazzi et al. 1999) observations.
The breadth of the line ($\sigma \sim 0.5$~keV; \ew $\sim \,
200-500~{\rm eV}$) if interpreted to be due to Doppler and
gravitational broadening is consistent with the presence of
relativistically deep gravitational potentials at the center regions
of AGN.  Recently, Wilms et al. (2002) reported with \XMM
observations of \mcg6 during a ``deep minimum'' (e.g. Iwasawa et al.
1996) an extremely broad line with breadth extending between
$\sim 3-7$~keV, confirming that the black hole is most probably
rapidly spinning.  See also Fabian et al. (2002) for results  on a
longer \xmm look.

In this Letter, we present the \chandra HETGS and simultaneous \rxte
 observations of \mcg6 in order to assess the narrow diskline
 component of the relativistically broadened Fe~K$\alpha$ emission and
 place limits on any intrinsically narrow core from distant
 (e.g. torus) material.
 Because the torus is very distant from the central black hole, we
 expect any line(s)  arising from this region to have velocity widths
 $<< \rm 2500 \rm \kmps$  FWHM (the width of the Balmer lines in
 \mcg6; Reynolds et al. 1997), and therefore not likely to be resolved
 by the HETGS ($\Delta \lambda \sim 1800\, \rm \kmps$ at the iron
 K$\alpha$ energies).  We also apply the same procedure to the data as
 was applied to the \asca data in T95, thereby clearly revealing the
 broad component in the \chandra data.

\section{Observations}
\label{sec:data}
MCG$-$6-30-15 was observed with the \chandra High Energy Transmission
Grating Spectrometer (HETGS; Canizares \etal 2002, in preparation)
from 2000 April 5--6, and again from 2000 August 21--22 (Fig.~\ref{fig-lc}). 
The total integration time was $\sim$~125~ks. The source varied by  $\sim 50\%$
during each observation. Simultaneous \rxte observations were made
between 2000 April 3--9, with  30~ks of usable data.

We reduce the spectral data from L1 (raw unfiltered event) files using
IDL processing scripts which are similar to the standard CIAO
processing. A more complete description is in Marshall, Canizares, \&
Schulz (2002). We restrict the event list to the nominal (0,2,3,4,6)
grade set, and remove event streaks in the S4 chip (these are not
related to the readout streak).  Event energies are corrected for
detector node-to-node gain variations.  The zeroth order position is
determined and the $\pm 1$ order events are extracted for source and
background regions. Bad columns are eliminated, as well as data which
are affected by detector gaps. The extracted MEG (HEG) events have
bins 0.01\AA\, (0.005\AA), or an ACIS pixel. The instrument effective
area is based on pre-flight calibration data.  We have assessed the
zeroth order pileup and find that the pileup fraction (defined to be
the fraction of all frames that have events  $> 2$ photons during that
frame) is 86\%.  (The average count rates are $\sim 1 \, \rm cts \,
s^{-1}$ from 0th order, and $1 \, \rm cts \, s^{-1}$ from the total of
all dispersed events.)  Accordingly, only $0.04 \, \rm cts \, s^{-1}$
are expected in single photon events making useful spectroscopic
information difficult to extract from zeroth order.
We use the software
package {\sc isis} (Houck \& DeNicola 2000) for the \chandra data
analysis

We extract PCA (Proportional Counter Array) light curves and spectra
from only the top Xenon layer using the {\sc ftools 5.1} software.
Data from layer~1 of PCUs 0, 2, and 4 are combined to improve
signal-to-noise at the expense of slightly blurring the spectral
resolution.   Good time intervals were selected to exclude any earth
or South Atlantic Anomaly (SAA) passage occulations, and to ensure
stable pointing.  We also filter out electron contamination events.
L7-240 background data are generated using {\sc pcabackest v4.0c}.
The PCA response matrix for the {\it RXTE } data set was created using
{\sc pcarsp  v7.11}.  Background models and response matrices are
representative of the most up-to-date PCA calibrations.

\vspace{0.2in}
\begin{figurehere}
\centerline{\psfig{file=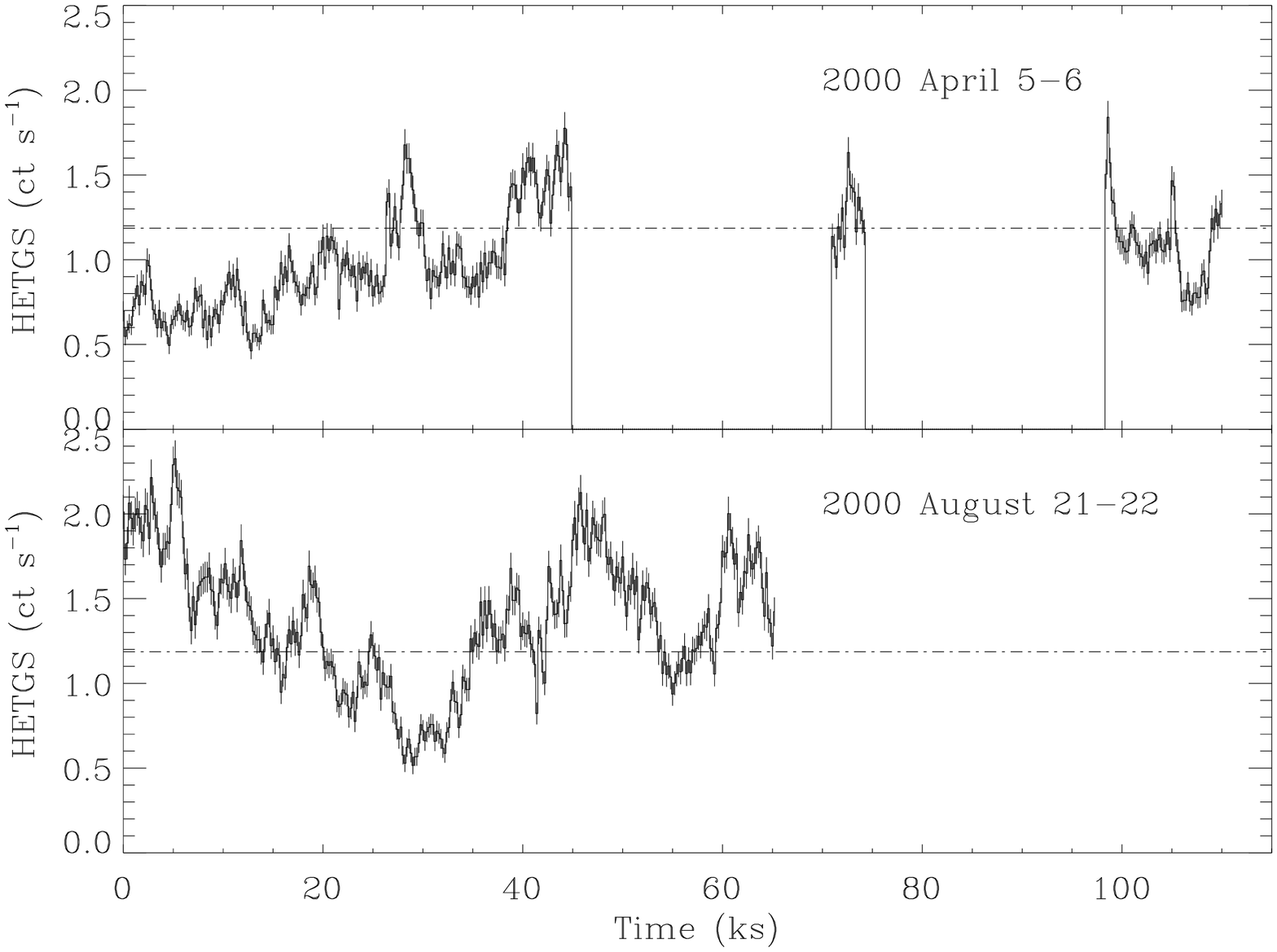,angle=90,width=8.5truecm,height=6.5truecm}}
\figcaption{\mcg6 light curve (excluding 0th order) binned at 200~seconds. 
  The dashed lines at $\sim 1.2\, \rm ct\, s^{-1}$ denote the mean
light curve count rate averaged over both observations.  The `high' and `low' states 
discussed in \S\ref{sec-narrowFe} are defined respectively to be above and below this 
value. 
\label{fig-lc}}
\end{figurehere}

\subsection{Spectral Results} \label{sec-cont}
The \rxte PCA is better suited for constraining the  continuum shape
as a consequence of its wide band coverage.  From these data, we
obtain the best-fit photon index  $\Gamma = 1.93 \pm 0.02$.  These
fits were performed in {\sc xspec} with a simple Galactic  ($4.06
\times 10^{20}\, \rm cm^{-2}$) absorbed power-law using the
3.3$-$4.6~keV and 8.0$-$10~keV spectral data ($\chi^2 / dof =
5.96/6$).  The $4.6-8$~keV energy range were excluded in order that we
may obtain the best description of $\Gamma$ with minimal bias from the
red wing of the broad iron line which can extend redwards to well
below 5~keV (e.g. Fig.~\ref{fig-hegasca},  Tanaka et al. 1995,
Iwasawa et al. 1996), or even farther (e.g. Wilms et al. 2002,
Fabian et al. 2002).

We next determined the
power-law normalization which best describes the \chandra HETGS
spectrum by fitting the MEG and HEG (respectively, medium and high
energy grating) $3-4$~keV data, with a power-law 
($\Gamma$ fixed at the best-fit \rxte value) modified by
$2 \times 10^{21} \, \rm cm^{-2}$ absorption (to accomodate 
the warm absorber which has $\approxlt 3$\% effect at 3~keV,
and no effect at $\approxgt 4$~keV). 
This limited energy range was chosen
in order to avoid contamination from the absorption features
attributed to the warm absorber below 3~keV (e.g. Lee et al. 2001,
Sako et al. 2002) and the broad Fe~K$\alpha$
emission above 4~keV.  The continuum is defined as the extrapolation
of this fit to include energies to 8~keV.  

\vspace{0.2in}
\begin{figurehere}
\vspace{0.2in}
\centerline{\psfig{file=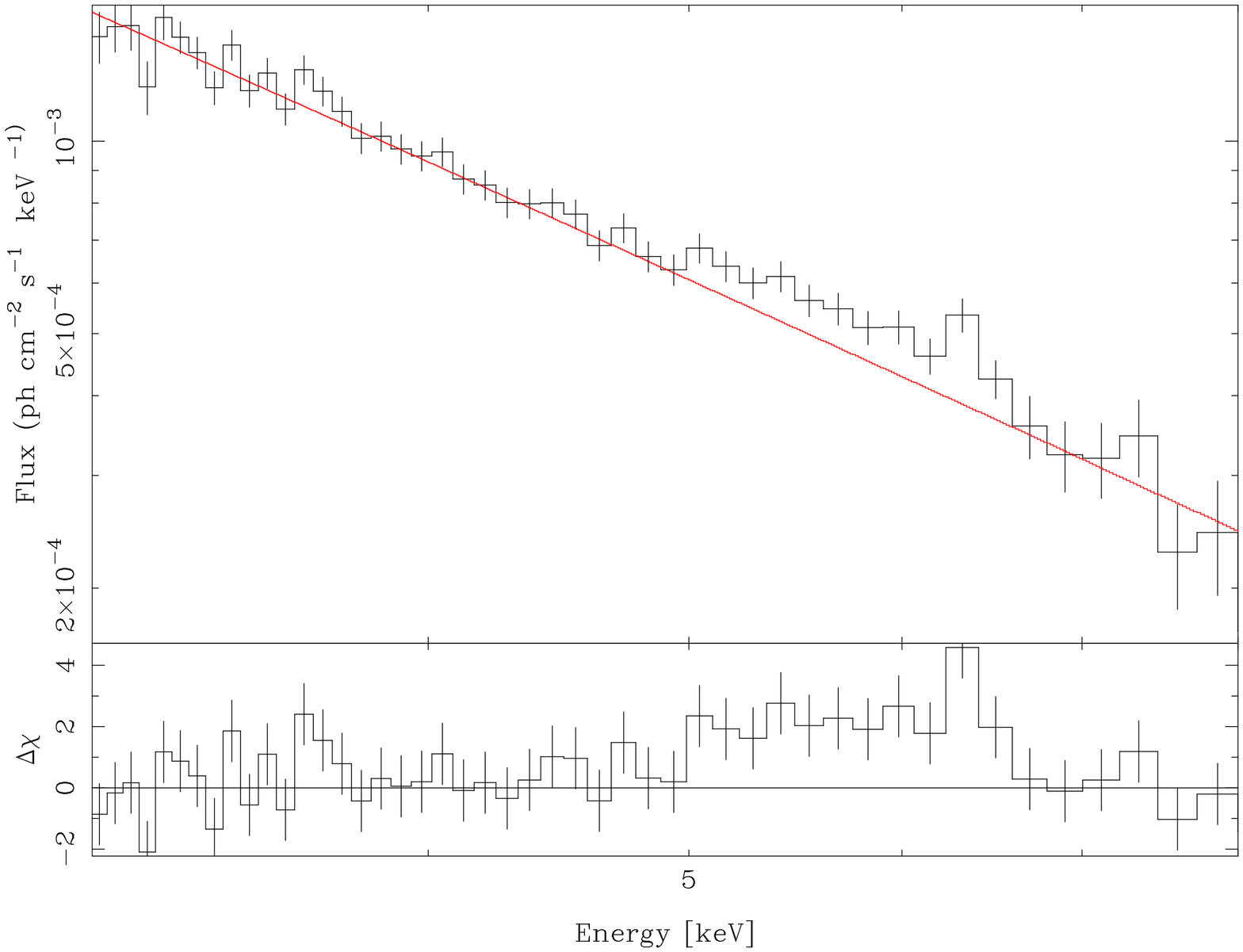,angle=270,width=8.5truecm,height=6.5truecm}}
\figcaption{The power-law continuum described in \S\ref{sec-cont} over-plotted on the HEG spectrum of 
\mcg6.  The data have been heavily binned (0.055~\AA\,)
in order to emphasize the excess emission
between $\sim 5-6.5$~keV that is attributed to the relativistically broadened 
Fe~K$\alpha$ emission. 
\label{fig-cont}}
\end{figurehere}
\vspace{0.2in}

\begin{figurehere}
\centerline{\psfig{file=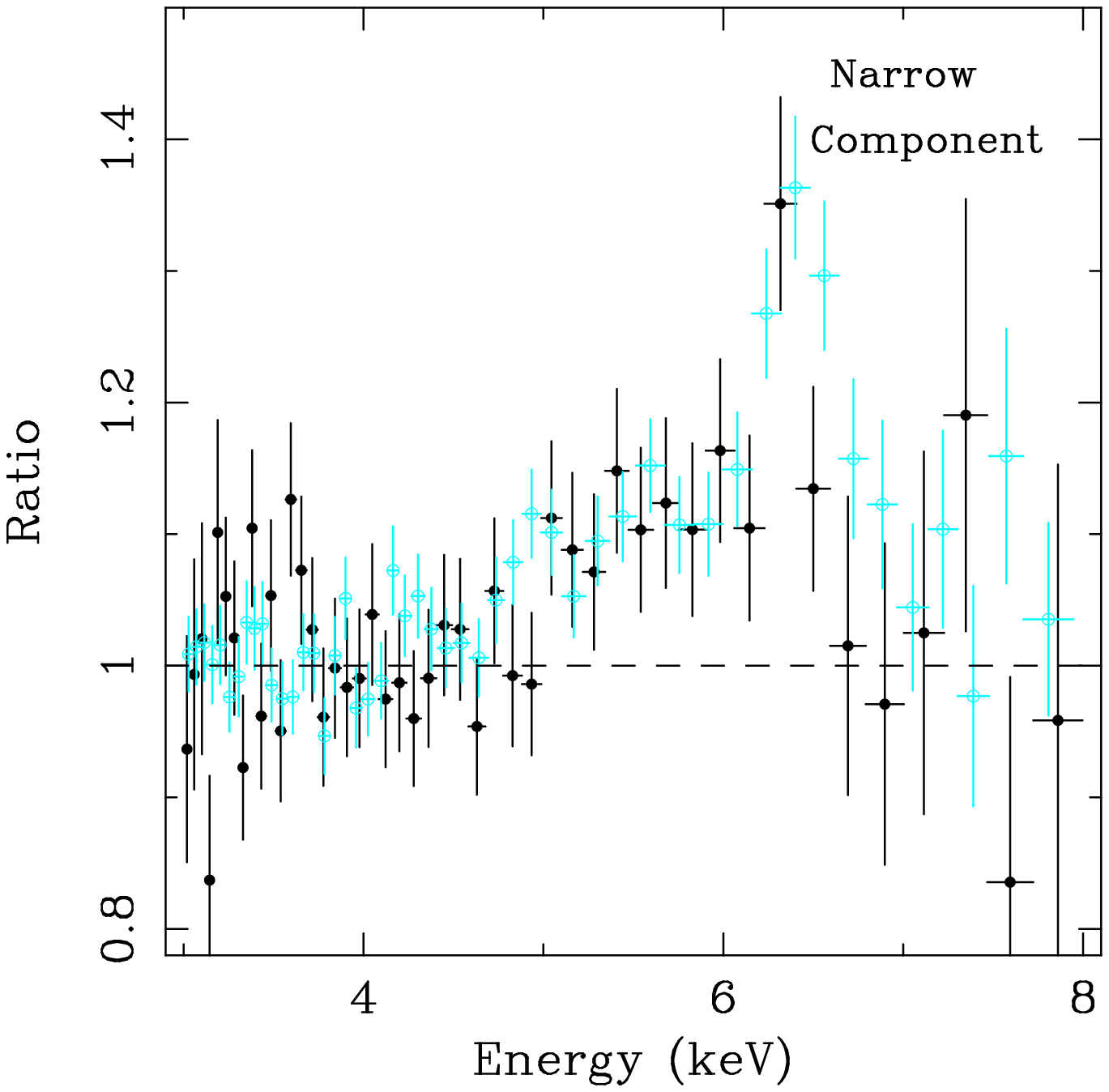,angle=270,width=8.5truecm,height=8.5truecm}}
\figcaption{The iron K$\alpha$ emission profile from the long (4.5 day) \asca observation
(Tanaka et al., 1995) in blue open circles, superposed on the heavily binned \chandra 
HEG spectrum (black filled circles) show that
the line is broad and skewed, extending far into the red.
\label{fig-hegasca}}
\end{figurehere}

\subsection{The relativistic diskline} \label{sec-diskline}
Fig.~\ref{fig-cont} illustrates the power-law continuum 
described previously over-plotted on the binned (0.055~\AA\,) 
HEG spectrum~$-$~notice the clear excess between 5 and 6.6~keV.  The
corresponding ratio plot of data$-$to$-$model in Fig.~\ref{fig-hegasca}
further shows that the heavily binned \chandra data reveals a line
profile very similar to that first discovered by T95. We note, however,
that since a simple power-law is used to derive the line profile seen
in this figure, it is possible that some fraction of the broad
excess at $\sim 5$~keV can have contributions from the reflection
spectrum known to exist in this source (e.g. Lee et al. 1998, 1999).

To further investigate the degree to which the relativistic disk line
can account for the line profile of Fig.~\ref{fig-hegasca}, we fit the
\chandra data with a power-law plus diskline model modified by
absorption.  The power-law component is that described previously and
the  diskline parameters are fixed at the T95 values : accretion disk
at inclination $\,i = 30^\circ$, respectively inner ($R_{\rm in} = 3.4
R_{\rm S}$) and outer ($R_{\rm out} = 10 R_{\rm S}$) radii , line
energy = 6.35~keV (6.4~keV in the galaxy frame), and radial emissivity
$\alpha = 3$ assuming a power-law-type emissivity function $\propto
R^{-\alpha}$ of the line.  This model gives $\chi^2 \, / d.o.f =
42/47$, and Fe~K$\alpha$ flux $(1.37 \pm 0.29) \times 10^{-4} \rm \,
ph \, cm^{-2} \, s^{-1}$, with equivalent width $W_{\rm K\alpha} \sim
295 \pm 118$~eV. This is in good agreement with the broad iron line
measured using  the \rxte data which has $W_{\rm K\alpha} = 398 \pm
58$~eV, comparable to previous \rxte (e.g. Lee et al. 1998, 1999) and
\asca measurements of this source (e.g. Iwasawa et al. 1996, 1999).
The apparent `sharp drop' at 6.7~keV (Fig.~\ref{fig-hegasca})  appears
resolved ($\Delta E > 70$~eV from peak to drop). 
A simple
Galactic absorbed power-law (as discussed previously) plus Gaussian
fit to the \rxte data gives $E_{\rm K\alpha(obs)} = 6.21 \pm
0.08$~keV,  Gaussian width $\sigma = 0.66 \pm 0.11$ (FWMH $\sim
73,000$~\kmps), and iron line flux  $I_{\rm K\alpha} = (2.13 \pm 0.31)
\times 10^{-4} \, \rm \ph \, cm^{-2} \, s^{-1}$.

\vspace{0.1in}
\begin{figurehere}
\centerline{\psfig{file=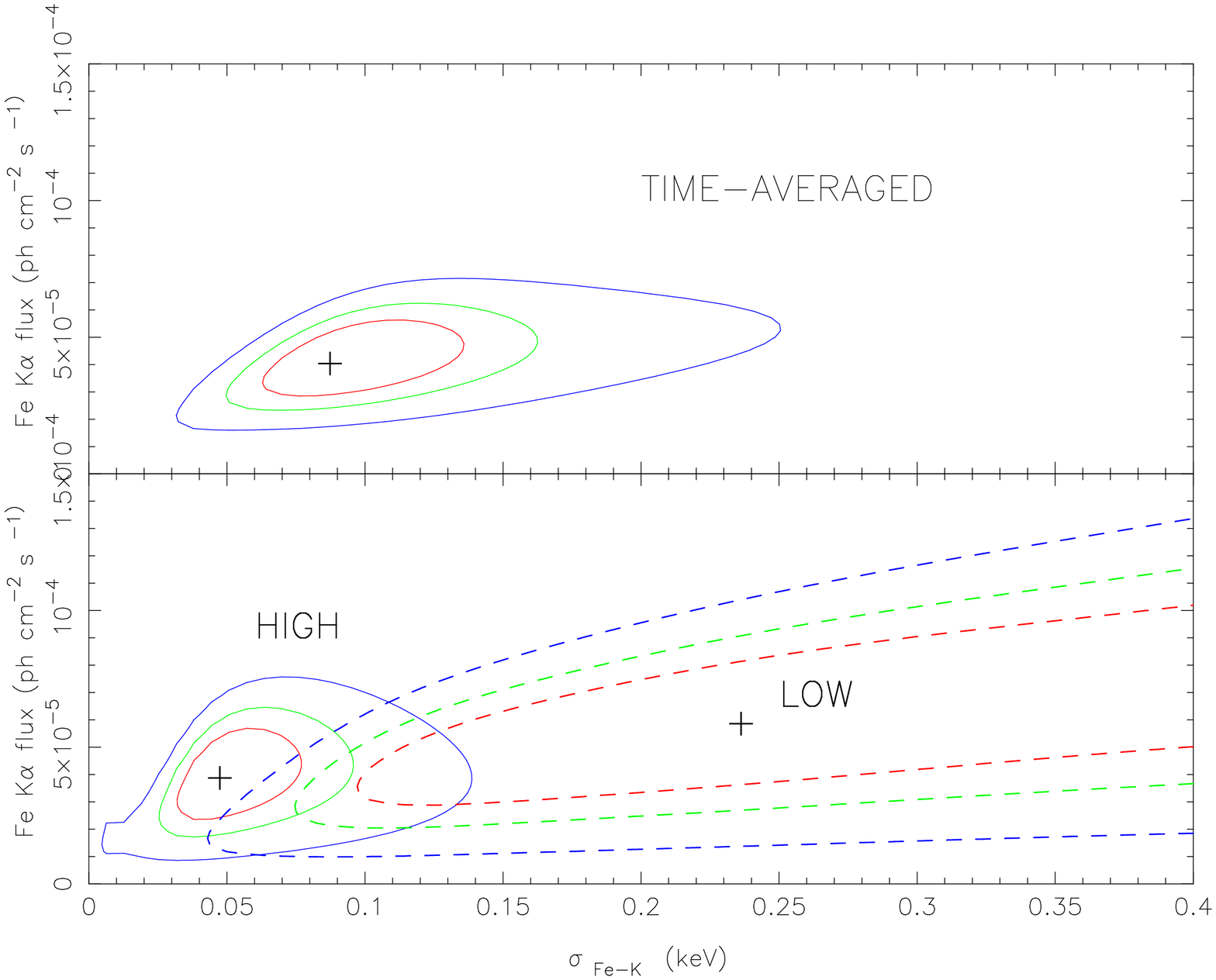,angle=270,width=8.5truecm,height=9.5truecm}}
\figcaption{Respectively from inner to outer contour, 68, 90, 99 per cent confidence limits 
on the iron K$\alpha$ narrow core.  The bottom panel shows the comparison of high (solid) versus low (dashed) state.
\label{fig-confcontours}}
\end{figurehere}
\vspace{0.2in}

\subsection{The Fe~K$\alpha$ narrow component} \label{sec-narrowFe}
We next address the formal detectability of the iron line narrow component
and resolution of its width.  
We note that the count rate of the full
125~ks observation at the iron line energies is $\sim 20-30 \rm \
cts\, bin^{-1}$ (where bin = 0.005\AA, or an ACIS pixel) and
therefore sufficient for statistically meaningful errors derived from
spectral fitting.  Table~1 shows the narrow component parameters when fitting
a Gaussian to the residuals of the best fit power law described in
\S\ref{sec-cont} (see also Fig.~\ref{fig-hilo}).
The 68, 90, and 99\%
confidence contours (3 free parameters and 515 degrees of freedom) 
of Fig.~\ref{fig-confcontours} show that 
the narrow component is resolved at 99\% confidence with a
full width at half maximum (FWHM)~$\sim 11,000 \, \rm km\,s^{-1}$
based on the full observation,  and $\sim 3600 \, \rm km\,s^{-1}$ from
the `high' continuum flux period  (see below).  
(The statistical significance of the excess counts in the narrow component
over a power law plus 10\% to account for the broad component is $> 99.9$\%.)
The line strengths
shown in Table~1 are consistent with  those found for the narrow core
by Iwasawa (1996, 1999).   Wilms et al. (2002) report an ``unresolved''
(at the \xmm EPIC resolution which is $\approxlt 4\times$ the \chandra HEG
at the iron energies) narrow iron line with \ew = 38~eV.~  
Fabian et al. (2002) attribute this to the blue wing of the disk line.

If there is a constant narrow core due to fluorescence from
material far from the black hole, such as a molecular torus, then it
should not vary during our observation and  be strongest during the
period of low continuum flux.  To test for the presence of an
intrinsically narrow core, we assess the variability nature of the narrow
component, by separating the data into a `high' (67~ks; 0.4-10~keV 
unabsorbed flux $f_{\rm high} = 5 \times 10^{-11} \, \rm erg\,cm^{-2}\,s^{-1}$) 
and `low' (58~ks; $f_{\rm low} = 3 \times 10^{-11} \, \rm erg\,cm^{-2}\,s^{-1}$)
  flux states arbitrarily defined as above and below the mean 
count rate (Fig.~\ref{fig-lc}) of the
time averaged 125~ks ($f_{\, \rm (0.4-10~keV)} = 
4 \times 10^{-11} \, \rm erg\,cm^{-2}\,s^{-1}$)
observation.  To ensure that there are sufficient
counts, these data were binned to 0.01~\AA\, to create the contour
plots corresponding to the `low' and  `high' state shown in
Fig.~\ref{fig-confcontours}.
It can be seen that the region of overlap between these states exists
only between the 90\% confidence contours. This would  indicate that
the joint probability  is $\sim 1$\% that the iron line narrow component
is a narrow core from distant material.  This is depicted in  Fig.~\ref{fig-hilo} which
shows that a comparison  of the high versus low flux data against the
best  fit power-law model reveals that the iron line narrow component is
effectively absent from the low flux state; $\sigma$ could  not be
well constrained during this state because a very  broad line (FMHM
$\approxgt 20,000 \, \rm km \, s^{-1}$) is required by the fit.  We
note the features blue-ward of  6.5~keV (e.g. at $\sim 6.82$~keV) seen
during the high  and low states are only marginally ($\approxlt 2
\sigma$) significant.

\vbox{
\footnotesize
\begin{center}

\begin{tabular}{lccc}
\multicolumn{3}{c}{\sc Table 1} \\
\multicolumn{3}{c}{\sc Parameters of the Fe~K$\alpha$ resolved narrow component} \\
\hline
\hline
{\em \rm Parameter } & {\em \rm $^a$ ALL} & {\em \rm {\sc $^b$ HIGH}} \\ \\
Exposure &  125 ks  & 67~ks   \\ \\
\hline
Restframe energy (keV) & $6.40^{+0.04}_{-0.05}$ &  $ 6.39^{+0.03}_{-0.02} $    \\ \\
Line flux ($10^{-5} \, \rm \ph \, cm^{-2} \, s^{-1}$)  & $4.2^{-1.6}_{+1.4}$  & $3.0^{+1.6}_{-1.5}$  \\ \\
Gaussian width $\sigma$ (keV)  &  $0.10^{+0.04}_{-0.04}$ & $0.03_{-0.02}^{+0.03}$  \\ \\
Velocity width ($\rm km \,s^{-1}$) &  $5000^{+2000}_{-2000}$ &  $1500^{+1400}_{-700}$ \\ \\
FWHM ($\rm km \,s^{-1}$) &  $11000^{+4600}_{-4700}$ &  $3600^{+3300}_{-2000}$ \\ \\
Equivalent width (eV)  & $110^{+38}_{-40}$ &  $62^{+33}_{-30}$  \\ \\
$\chi^2$ / dof  &  562/ 515 & 586 / 515   \\

\hline
\end{tabular}

\parbox{3.2in}{
\vspace{0.1in}
\small\baselineskip 9pt
\footnotesize
\indent
Fits are assessed using the HEG data and errors are quoted at 90\% confidence. 
$^a$ Data bins are 0.005\AA\,.  Note that we were 
unable to constrain the velocity width of the narrow component during the low state. \\
}
\label{tab-edge}
\end{center}
\normalsize
}

Using the 125~ks observation, we can set limits on the  unresolved
core by fixing the line energy at 6.4~keV and $\sigma$ to the
resolution of the HEG ($\sim 1800 \, \kmps$~FWHM) at that energy,  and
assessing the 90\% confidence for the line flux.  The upper limit on
an intrinsically narrow core is estimated to be $F_{\rm K\alpha} \sim
6 \times 10^{-6} \rm \ ph\,cm^{-2}\,s^{-1}$, fractionally $<$~15\% of
the line strength for the narrow component shown in Table~1.

\vspace{0.2in}
\begin{figurehere}
\centerline{\psfig{file=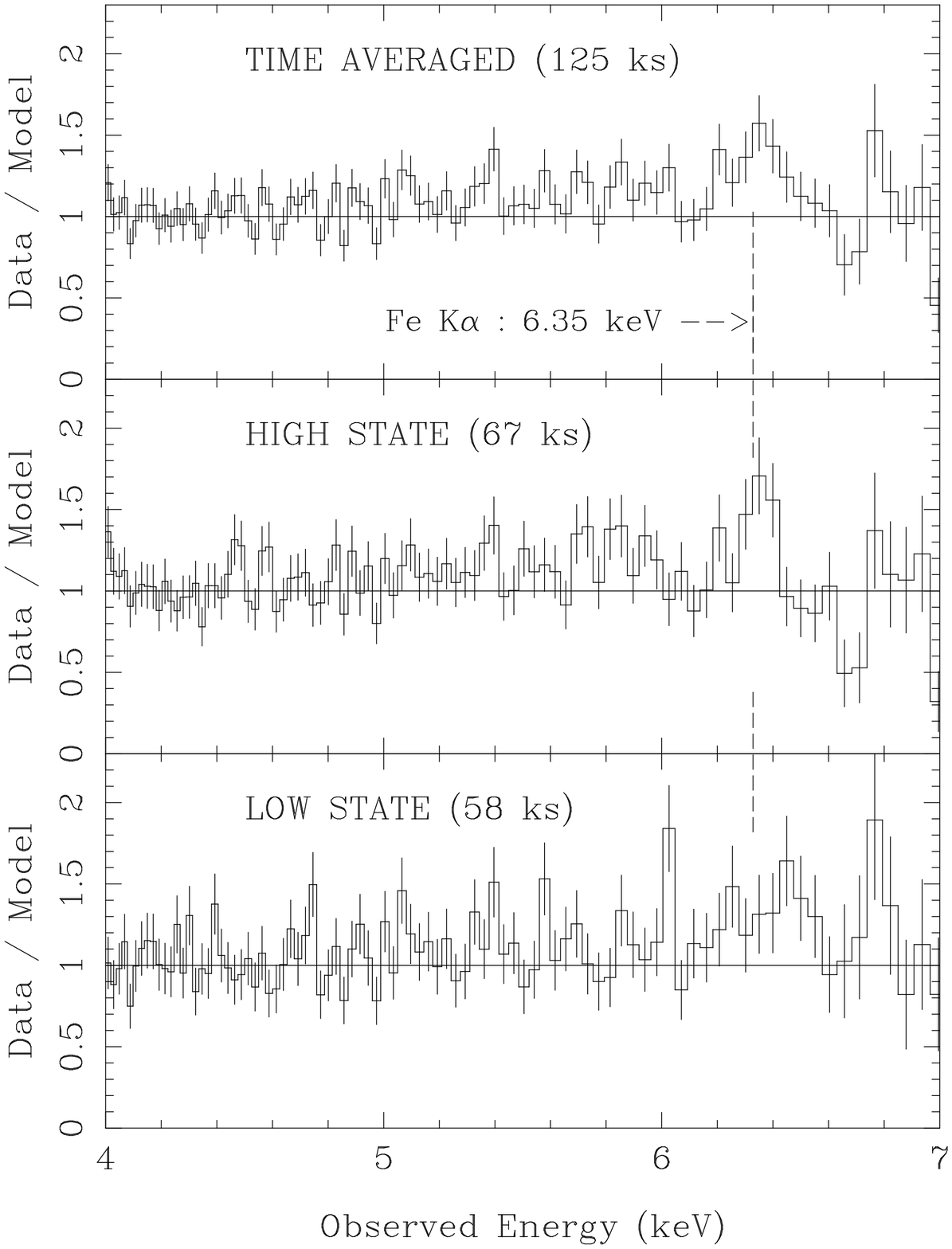,angle=0,width=8.5truecm,height=9.5truecm}}
\figcaption{Ratio of data : power-law model shows that the
Fe~K$\alpha$ narrow core is more prominent during the high state.
Note that 6.35~keV is the  observed energy of the 6.4~keV Fe~K$\alpha$
fluorescent emission in the rest  frame of \mcg6.  Data bins are
0.015\AA.
\label{fig-hilo}}
\end{figurehere}
\vspace{0.2in}

\section{Discussion}
A clear measurement of the iron line narrow component with \chandra has
implications for resolving differences in the interpretation of the
breadth and origin of the iron line in \mcg6. This pertains
specifically to whether part of the iron line originates from a torus
outside the broad line region (Ghisellini, Haardt \& Matt 1994; Yaqoob
et al. 2001, Reeves et al. 2001), as opposed to the blue horn of the relativistic diskline.

The possibility is small that a significant fraction of the broad line
in \mcg6 is due to some distant material (e.g. a torus) based on the
\chandra HEG data. It is more likely that the resolved narrow
component is  consistent with being the blue peak of a diskline (or
possibly emission  from the outer radii of the concave disk proposed
by Blackman 1999). In particular, the \chandra HETGS data resolves the
narrow component of the \feka emission at 6.4~keV (6.35~keV observed)
with   a velocity width $\sim 4700 \, \rm km\,s^{-1}$ (FWHM~$\sim
11,000 \, \kmps$). It has an \ew $\sim 110$~eV, and is most pronounced
when the continuum flux is high (FWHM~$\sim 4000 \, \kmps$).

\begin{itemize}
\item
The resolved narrow component of the \feka emission seen in the HEG data 
of \mcg6 is significantly broader than expected from a torus
outside the BLR (e.g. as seen in NGC~5548 with the HETGS -- Yaqoob et al. 2001,
or Markarian 205 with \xmm EPIC -- Reeves et al. 2001).

\item Since the torus is thought to be $\approxgt$~1~pc from the
central  source, reflection from this region should remain constant at
average intensity levels.  Therefore, we expect that  such a component
would be especially pronounced when the continuum flux is low.  Yet,
the narrow component is virtually absent from the low flux state in our
data.  This is consistent with the findings from \asca data (Iwasawa
et al. 1996, 1999).

\item The 90\% confidence limit on an intrinsically narrow core is
$F_{\rm K\alpha} \sim 6 \times 10^{-6} \rm \, ph\,cm^{-2}\,s^{-1}$,
fractionally $<$~15\% of the line strength of the narrow component, or
$\sim 16$~eV.  This is significantly less than the $\sim 60$~eV  found
for the intrinsically narrow line thought to be from  distant material
found in the Seyfert~1.9 MCG$-$5-23-16  (Weaver et al. 1997).

\item The difference in the iron line between  the periods of low
versus high continuum flux is seen in the width rather than the flux in
the line.  This implies that the line cannot come from
very distant material but that it comes from kinematically different
(or kinematically variable) material in the low and high states.

\end{itemize}


In summary, the \chandra HETGS data of MCG--6-30-15 confirm the
presence of a broad line, resembling in shape and strength the line
seen first with \asca (T95). The narrow component at $\sim 6.4$~keV is
not narrow enough to be from distant matter and is plausibly the blue
wing of a relativistic line. Any intrinsically narrow core has an
equivalent width of less than 16~eV, a small fraction of the 
total resolved narrow component (see Table~1). This is less than 25\% of that
expected if the X-ray source illuminates a distant molecular torus as
commonly expected from Seyfert galaxy unification models (see e.g.
Ghisellini et al. 1994 and references therein).

\section*{acknowledgements}
We thank Paul Nandra for useful comments.
The work at MIT was funded in part by contract SAO SV1-61010 and NASA
contract NAS8-39073.  ACF thanks the Royal Society for support. KI
thanks PPARC for support.

\end{document}